\documentclass[twocolumn,showpacs,preprintnumbers,amsmath,amssymb,aps,prb]{revtex4}
\usepackage{graphicx}

\pdfoutput=1
\usepackage{ifpdf}
\usepackage{dcolumn}
\usepackage{bm}
\usepackage{mathrsfs}
\usepackage{amsmath}
\usepackage{amsfonts}
\usepackage{amssymb}
\usepackage{mathrsfs}
\usepackage{color}
\usepackage{multirow}
\usepackage[import]{xy}
\newcommand{\abs}[1]{\left| #1 \right|}

\begin{document}
\tolerance=1
\emergencystretch=\maxdimen
\hyphenpenalty=10000
\title{Dynamics of self-organized driven particles with competing range interaction}

\author{H.J.~Zhao}
\affiliation{
Department of Physics, University of Antwerpen, Groenenborgerlaan 171, B-2020 Antwerpen,
Belgium
}
\author{V.R.~Misko}
\affiliation{
Department of Physics, University of Antwerpen, Groenenborgerlaan 171, B-2020 Antwerpen,
Belgium
}
\author{F.M.~Peeters}
\affiliation{
Department of Physics, University of Antwerpen, Groenenborgerlaan 171, B-2020 Antwerpen,
Belgium
}

\date{\today}

\begin{abstract}
Non-equilibrium self-organized patterns 
formed by particles interacting through competing range interaction are 
driven over a substrate by an external force. 
We show that, with increasing driving force, the pre-existed static patterns evolve into dynamic patterns either via disordered phase or depinned patterns, or via the formation of non-equilibrium stripes. 
Strikingly, the stripes are formed either in the direction of the driving force or in the transverse direction, depending on the pinning strength. 
The revealed dynamical patterns are summarized in a dynamical phase diagram. 
\end{abstract}

\pacs{89.75.Kd,82.70.Dd,74.25.Uv}
\maketitle

\section{Introduction} 

Most of the studies on self-assembly focused on equilibrium systems which form highly ordered arrays (crystals). 
Equilibrium systems with a competing repulsive and attractive interaction~\cite{Seul1995,Reichhardt2003,Reichhardt2003a,Nelissen2005,Vedmedenko2007,Zhao2012a}
demonstrate a variety of morphologies with an inhomogeneous distribution of particles, including stripes, clusters, bubbles, etc.
However, 
most important are {\it dynamic} self-assembling systems, i.e., out-of-equilibrium systems that form their characteristic order when dissipating energy~\cite{Whitesides2002,Ball1999}.
Understanding and controlling such systems is required in order to maximize the value of self-assembly as a strategy for synthesis and fabrication and, e.g., for the understanding of the role of self-assembly in biology.

Interacting systems driven on a substracte include, e.g., 
colloids~\cite
{Reichhardt2002a,Cao2003,Pertsinidis2008,Bohlein2012},
charge density waves~\cite{Gruner1988,Balents1995}, 
Wigner crystals~\cite{Reichhardt2001a}, 
and 
vortex matter in superconductors~\cite
{Bhattacharya1993,Yaron1995,Duarte1996,Pardo1996,Pardo1998, 
Koshelev1994,Moon1996,Giamarchi1996,Balents1997,Olson1998,Rosenstein2010,Pogosov2010}. 
Thus, in a two-dimensional (2D) colloidal crystal driven across ordered substrates, kinks and antikinks were experimentally observed~\cite{Bohlein2012}. 
For vortex matter driven on a random substrate, 
various dynamical regimes: pinned, plastic, and elastic flow 
were revealed 
in experiments~\cite{Bhattacharya1993,Yaron1995,Duarte1996,Pardo1996,Pardo1998} 
and in theory~\cite{Koshelev1994,Moon1996,Giamarchi1996,Balents1997,Olson1998,Rosenstein2010,Pogosov2010}. 
Using a periodic substrate, instead of random, results in rich dynamics including negative differential resistivity 
of either N-type~\cite{Reichhardt1997}
or N- and S-type~\cite{Misko2006b,Misko2007a}.
Furthermore, driven vortex matter on quasiperiodic substrates~\cite{Misko2005,Misko2006,Villegas2006,Kemmler2006,Silhanek2006} revealed electron-like or hole-like ``vortex conductivity''~\cite{Misko2010}.
In the above examples, the interaction between the fundamental units (i.e., colloids or vortices) was purely repulsive with a single length scale.

In case of competing interaction
characterizied by two (or more) length scales, 
systems self-organize forming equilibrium, or at least metastable, patterns.
The main question is, how the pre-existed static patterns will evolve being driven on a random or regular substrate? 
And, will they evolve in some other, dynamically-induced, non-equilibrium patterns which do not exist in equilibrium?
These are still open important questions for the general case of competing interaction and for an arbitrary potential-energy profile of the substrate. 
Only a few studies have been performed for some special cases. 
For example, the dynamical reordering in the plastic regime~\cite{Reichhardt2003a,Reichhardt2003,Xu2011b,OlsonReichhardt2011}
was examined in a system with competing short and long-range
interactions~\cite{Reichhardt2003,Reichhardt2003a}.

In the previous studies, the interaction at long distance was repulsive, since an attractive interaction leads 
to phase separation.
However, 
as shown recently~\cite{Zhao2012a}, 
a competing interaction with a repulsive core and a fast decaying attractive tail can induce long-living metastable states.
This model~\cite{Zhao2012a} 
is applicable to a number of 
diverse 
physical systems with fast decaying interaction, e.g., atoms and molecules (Lennard-Jones), colloids and ferrofluids. 
Also, it can model inter-vortex interaction in low-$\kappa$
~\cite{Brandt2011,Xu2011,Xu2011b} 
and multi-band superconductors such as $MgB_2$, 
where inhomogeneous vortex patterns, e.g., gossamers and stripes, were 
recently 
revealed experimentally~\cite{Moshchalkov2009,Nishio2010}.
Those unusual patterns result from a non-monotonic inter-vortex interaction as argued in 
~\cite{Babaev2005,Babaev2010,Chaves2011,Komendova2011,Lin2011,Komendova2012,Silaev2012}.
Furthermore, recent scanning Hall probe microscopy 
measurements~\cite{Gutierrez2012} revealed stripe vortex patterns in $MgB_2$ which were not due to inhomogeneous pinning but rather suggestive of another source of ordering, e.g., an external driving bias that might orient the stripes. 
Therefore, studying the dynamics of systems with competing range interaction is both of fundamental interest and 
useful for the understanding of unusual experimental patterns.

\section{Model} 

We study the dynamics of particles with a non-monotonic interaction moving in 2D, by numerically integrating the Langevin equations using molecular-dynamics (MD) simulations. 
The  overdamped equation of motion is~\cite{Reichhardt2003,Reichhardt2003a,Reichhardt2005,Zhao2012a,Zhao2012}:
\begin{equation}
\label{Md}
\eta \textbf{v}_i=\textbf{F}_i=\sum_{j\neq i} \textbf{F}_{ij}+\textbf{F}_{i}^p+\textbf{F}_{i}^T+\textbf{f}_d.
\end{equation} 
We follow the model introduced in~\cite{Zhao2012a,Zhao2012} and used in~\cite{Reichhardt2012a}, 
where the inter-particle interaction 
$\textbf{F}_{ij}$
has a repulsive core and an attractive tail,
\begin{equation}
\label{eq-interaction_1}
\textbf{F}_{ij}=F_0\bigg(\frac{K_1(r_c/\lambda)}{K_1(br_c/\lambda)}K_1(br_{ij}/\lambda)-K_1(r_{ij}/\lambda)\bigg)\hat{\textbf{r}}_{ij}, 
\end{equation}
where 
$K_{1}$ is the first-order modified Bessel function and 
$F_0$ ($\lambda$) is the unit of force (length). 
$\textbf{F}_{i}^p$
is the interaction of the particles with pinning sites: 
\begin{equation} 
\label{eq-fp} 
\textbf{F}_i^{p}=\sum_k^{N_p}
\left({f_p}/{r_p}\right)
\abs{\textbf{r}_i-\textbf{r}_k^{(p)}}
\Theta
\left({r_p-\abs{\textbf{r}_i-\textbf{r}_k^{(p)}}}/{\lambda}\right)
\textbf{\^{r}}_{ik}^{(p)}. 
\end{equation}
Here $N_p$ is the number of pinning sites, $f_p$ is the maximum pinning force, 
$r_p$ is the pinning range
$\Theta$ is the Heaviside step function, and
$\textbf{\^{r}}_{ik}^{(p)}=(\textbf{r}_i-\textbf{r}_k^{(p)})/\abs{\textbf{r}_i-\textbf{r}_k^{(p)}}$.
$\textbf{F}_{i}^T$ 
is the thermal stochastic force, which obeys the following conditions: 
\begin{equation} 
\langle F_i^T(t)\rangle =0 
\end{equation}
and 
\begin{equation} 
\langle F_i^T(t) F_j^T(t^\prime)\rangle
= 2\eta k_B
T\delta_{ij}\delta(t-t^\prime). 
\end{equation}
The last term 
$f_d$ 
in Eq.~(\ref{Md}) 
is the applied driving force. 
We consider a 2D square simulation box $L_{x} \times L_{y}$ 
(where $L_{x} = L_{y} = L= 120$) 
in the $xy$-plane and apply periodic boundary conditions~\cite{pbc}. 
For the interaction force given by Eq.~(\ref{eq-interaction_1}),
which decays exponentially for large $r$, 
we use a cut-off~\cite{cutoff} $r=8$. 
Thus, the interaction in our model is {\it finite-range}. 
The initial states are obtained by performing simulated annealing simulations 
of interacting particles without driving. 
Then, the driving force is turned on at zero temperature~\cite{temp}.

\section{Dynamical regimes} 

The dynamics of the system is determined by three main factors: the inter-particle, particle-pinning interactions, and the applied driving. 
To characterize the dynamical regimes, we calculate the average velocity $<v>$ versus $f_d$ for varying 
$f_p$. 
In general, 
for very weak $f_d$, most of the particles are pinned resulting in 
$<v> \gtrsim 0$. 
In the opposite limit of large $f_d$ ($>f_p$), all the particles are depinned, 
and 
$<v> \propto f_d$. 
The intermediate range of $f_d$ ($0<f_d<f_p$) exhibits rich dynamics. 
Depending on $f_p$, $<v>-f_d$ curves acquire well-distinguished parts
(Fig.~\ref{fig:fv}) 
indicative of different dynamical regimes. 
In the Supplemental Material (Sec.~VIII), examples of snapshots are presented illustrating various dynamical regimes and transitions (crossovers) between them. 

Below we present our results for specific pinning regimes and the following interaction parameters: $r_c=2.4$ and $b=1.1$, the particle density $n=N/L^2=0.0139$, where $N=2000$ is the total number of particles in the simulation cell, 
and the density of random pinning $n_p=n$. 
Note that static configurations for these parameters are labyrinths~\cite{Zhao2012a}. 

\begin{figure}[t]
  \includegraphics[width=0.45\textwidth]{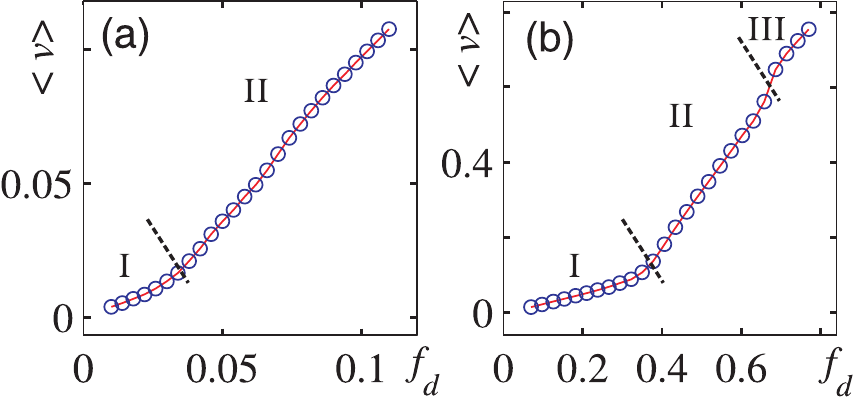} 
  \caption{The average velocity $<v>$ versus the driving force $f_d$ for different strength of pinning force $f_p=$ (a) 0.1; (b) 0.7. For weak pinning, there are two dynamical regimes. For stronger pinning (b), the curve has an ``S-like'' shape, which exhibits three dynamical regimes.}
\label{fig:fv}
\end{figure}

\begin{figure}[t]
  \includegraphics[width=0.45\textwidth]{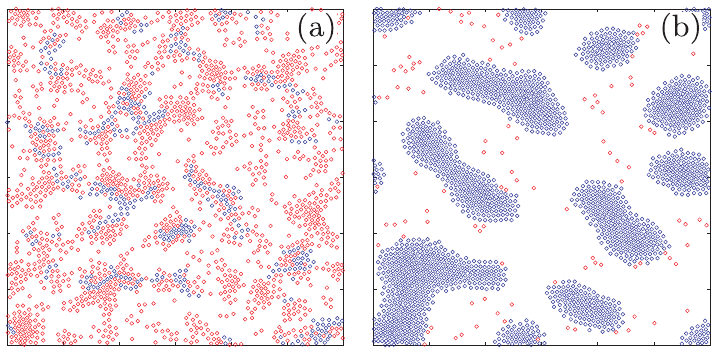} 
    \caption{Dynamical patterns for $f_p=0.1$ and $f_d=$ (a) 0.014; (b) 0.07. The mobile (pinned) particles are shown by blue (red) dots. As $f_d$ increases, the dynamical pattern changes from partially pinned labyrinths (a) to moving clusters (b). 
}
\label{fig:fp_small}
\end{figure}

\subsection{Weak pinning} 

Very weak pinning 
with strength 
($f_p\sim0.1$), 
which is 
comparable to the elastic force due to the inter-particle interaction, 
is insufficient to destroy the static morphology. 
Therefore, for weak driving ($f_d<0.03$), particles only flow {\it inside} the labyrinths 
(Fig.~\ref{fig:fp_small}(a)) 
(these are unpinned partcles which are trapped by the pinned neighbors, i.e., ``collectively pinned'') 
resulting in very small $<v>$ 
(Fig.~\ref{fig:fv}(a)). 
For $f_d$ larger than some threshold (e.g., $f_d=0.038$), 
most of the particles depin, although $f_d \ll f_p$. 
This {\it collective} depinning occurs as follows: parts of the labyrinths depin and form small clusters 
which  
repin again 
if driving is weak. 
For stronger driving, they merge with pinned parts and depin them, leaving only a few individually pinned particles (see Fig.~\ref{fig:fp_small}(b)). 
As a result, the number of mobile particles increases with time, until most of them are depinned.

\begin{figure}[t]
  \includegraphics[width=0.3\textwidth]{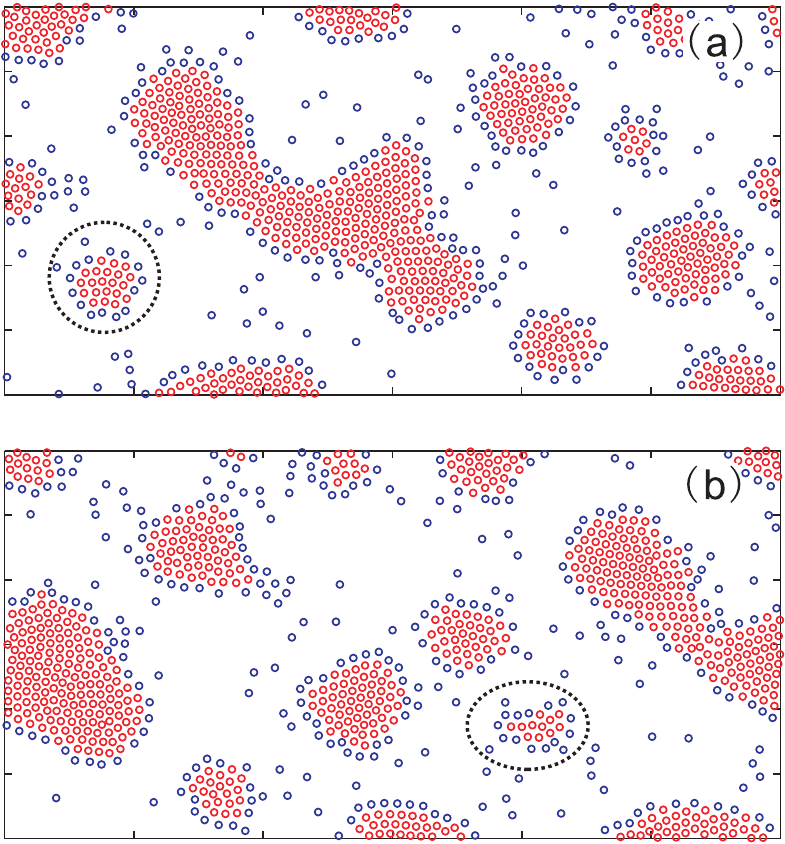} 
  \caption{Evaporation of a small cluster (marked by black dotted line) following a large cluster (moving from the left to the right) for $f_d=0.062$, $f_p=0.1$, and (a) $t=1900\Delta t$; (b) $2500\Delta t$. 
}\label{fig:evaporation}
\end{figure}

\subsubsection{The optimal cluster size} 
 
Merging of small clusters is not the only mechanism of increasing the cluster size. 
Strikingly, they can also ``grow'' by collecting individual particles generated by ``evaporating'' small clusters.

To understand this effect, we explore the analogy between a cluster moving on a pinning landscape 
and a droplet of water moving on a 
prewetted substrate~\cite{Gennes1985,Bonn2009}. 
The size of the droplet increases if the 
substrate is over-wetted, and vice versa: the size of the droplet decreases if the substrate is dry. 
Therefore, there exists a critical wettability when the droplet keeps its size.
Translating this to the language of our system, 
larger clusters have larger attraction to their tail particles. 
Therefore, the critical ``wettability'' of large clusters is smaller than that of small ones. 
Now, considering a small cluster following a large one, the substrate passed by the large cluster is always {\it dry} for the small cluster.
Thus, the small cluster {\it evaporates} and leaves an over-wetted substrate for the large clusters.

Fig.~\ref{fig:evaporation} illustrates the evolution of a small cluster 
following a large cluster. 
The small cluster remains stable until the cluster in front of 
it 
becomes large enough (Fig. \ref{fig:evaporation}(a)). 
Then the small cluster gradually loses its particles 
(see Fig.~\ref{fig:evaporation}(b)), and finally 
disappears. 
As a result, a steady state is formed when all the clusters are large and well separated.

\begin{figure}[t]
  \includegraphics[width=0.45\textwidth]{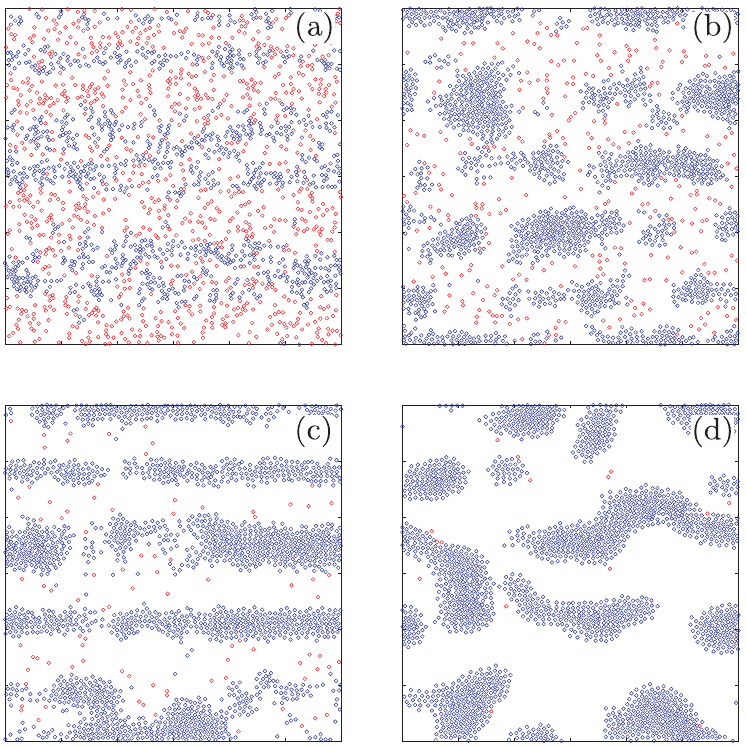} 
    \caption{Dynamical patterns for 
$f_p=0.5$ and 
$f_d=$ (a) 0.21; (b) 0.39; (c) 0.43; (d) 0.49. 
Red (blue) dots stand for mobile (pinned) particles. 
}\label{fp_intermediate}
\end{figure}

\subsection{Intermediate pinning} 

For 
the pinning force 
$f_p$ larger than the 
strength of the 
typical inter-particle interaction force ($0.3\leq f_p \leq 0.7$), the pre-existed patterns are destroyed. 
There are three dynamical regimes in case of intermediate pinning strength (Fig.~\ref{fig:fv}(b)). 
In regime I, the pinning force is larger than the elastic force due to the inter-particle interaction, while the driving force is smaller. 
Therefore, part of the labyrinths can still exist, i.e., in the form of deformed small clusters. 

A stronger 
driving force 
$f_{d}$ 
can overcome the elastic force between particles and the pinning force, resulting in more mobile particles, which 
increases $<v>$. 
Further increase of $f_d$ increases the density of mobile particles, until all the particles are depinned (regime II). 
In turn, 
this regime can be divided into three sub-regimes: 
(i) 
$f_d$ is just above the elastic force: 
the resulted dynamical patterns are disordered 
(Fig.~\ref{fp_intermediate}(a)); 
(ii) 
stronger $f_d$: the mobile particles form clusters 
(Fig.~\ref{fp_intermediate}(b)) 
which elongate in the direction of driving 
(i.e., for $r_c>2.3$, see Ref.~\cite{Zhao2012a}) 
when the ``head" and ``tail'' of a cluster move with different velocity because of the different ``friction'' due to the pinning 
(note that the motion in the medium, without pinning, is overdamped, while the additional ``friction'' is related to the pinning); 
(iii)
even larger $f_d$: the elongated clusters connect to each other, which restores a continuous particle flow 
(Fig.~\ref{fp_intermediate}(c)). 

Although the stripe formation has been found in many physical systems, e.g., driven vortices 
\cite{Xu2011,Xu2011b,Zhao2012a,Zhao2012}, 
2D 
electrons~\cite{Reichhardt2003,Reichhardt2003a,Reichhardt2005} and systems with shoulder potential~\cite{Malescio2003}, 
the stripe formation and orientation in our case is different. 
In those studies, the inter-particle interaction for long distance was repulsive, which prevented the stripes from merging and arranged them parallel. 
In contrast, here the weak attractive tail leads to the accumulation of single mobile particles but is not sufficient for merging the stripes locked by the random pinning. 
As a consequence, the stripes have varying width and inter-stripe distance, in contrast to the ones resulting from a purely repulsive tail
~\cite{Reichhardt2003}. 

Finally, for $f_d>f_p$, all the particles are depinned, 
the clusters or stripes are not oriented 
(Fig.~\ref{fp_intermediate}(d)), 
and $<v>$ increases linearly versus $f_d$ 
(regime III)~\cite{Reichhardt1997,Misko2006b,Misko2007a}.

\begin{figure}[t]
  \includegraphics[width=0.45\textwidth]{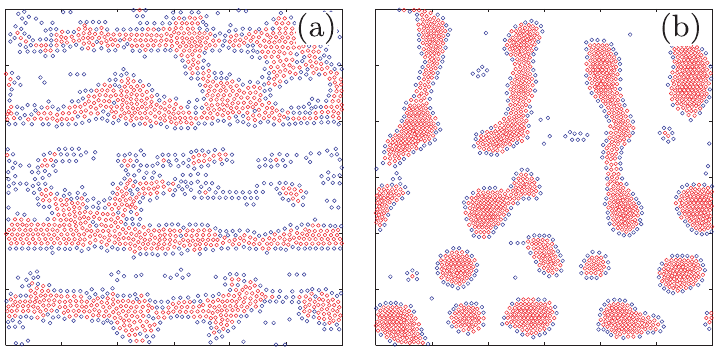} 
    \caption{Stripes in fast sliding dynamical regimes. (a) longitudinal stripes for $r_c=2.5$, $f_p=1.7$, and $f_d=1.8275$; (b) transverse stripes for $r_c=2.3$, $f_p=1.7$, and $f_d=2.465$.}\label{fp_even_strong}
\end{figure}

\subsection{Strong pinning: Formation of transverse stripes} 
 
For sufficiently strong pinning ($f_p\geq0.9$), even a small number of pinning sites destroy the pattern structure. 
As a result, for a broad range of $f_d$, the dynamical patterns are disordered.
Moving clusters are formed when $f_{d}$ 
is very close to the depinning threshold: 
$f_{d} \lesssim f_{p}$. 

For the so-called ``fast sliding regime'', when $f_d>f_p$, the random pinning has the same impact on the moving patterns as the thermal Langevin force~\cite{Koshelev1994}. 
Strong pinning orients the sliding triangular lattice in the longitudinal direction 
(Fig.~\ref{fp_even_strong}(a)), 
i.e., particles move in {\it static channels}
which were found for moving elastic lattices with disorder 
~\cite{Giamarchi1996}. 
Once these channels are formed, there are energy barriers 
for transverse motion, i.e., the motion along the longitudinal channels is {\it locked}. 
These transverse energy barriers can be overcome by a weak bias force, e.g., a weak external driving in the transverse 
direction.

We use the paradigm of static channels to explain 
a very unusual behavior revealed for $r_c \lesssim 2.3$: 
the formation of 
striking {\it transversal} stripes 
(see Fig.~\ref{fp_even_strong}(b)). 
To understand this unusual behavior, 
recall that in the absence of pinning, 
particles try to form, due to the attraction force, 
circular clusters for this value of  
$r_{c}$~\cite{Zhao2012a}. 
The aggregation due to the inter-particle attraction, 
in principle,  
overcomes the elongation of clusters due to the formation 
of static channels. 
However, 
the latter process is much {\it faster} than the former one, and as a result, 
for strong enough driving,  
the clusters turn out to be {\it dynamically locked} by the static channels. 
Merging clusters still tend to acquire circular shape, but due to the existing transverse barriers, they become ``stretched'' 
along the transverse direction, 
thus resulting in the observed 
transverse stripes~(Fig.~\ref{fp_even_strong}(b)).

\begin{figure}[tt]
  \includegraphics[width=0.45\textwidth]{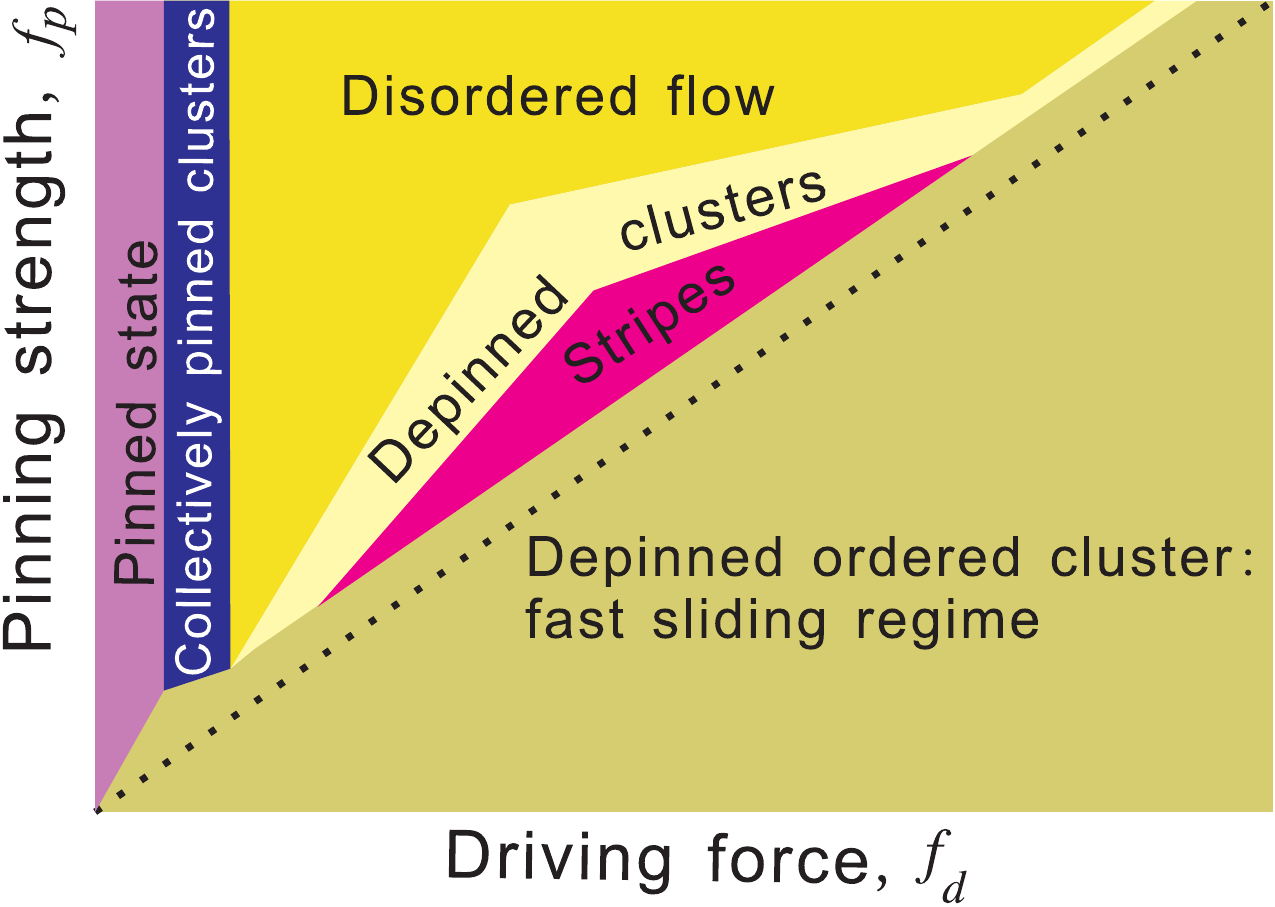} 
  \caption{Schematic dynamical phase diagram in the plane 
$f_p$ versus $f_d$. 
}\label{fig:phase}
\end{figure}

\begin{figure}
  \centering
  \includegraphics[width=\columnwidth]{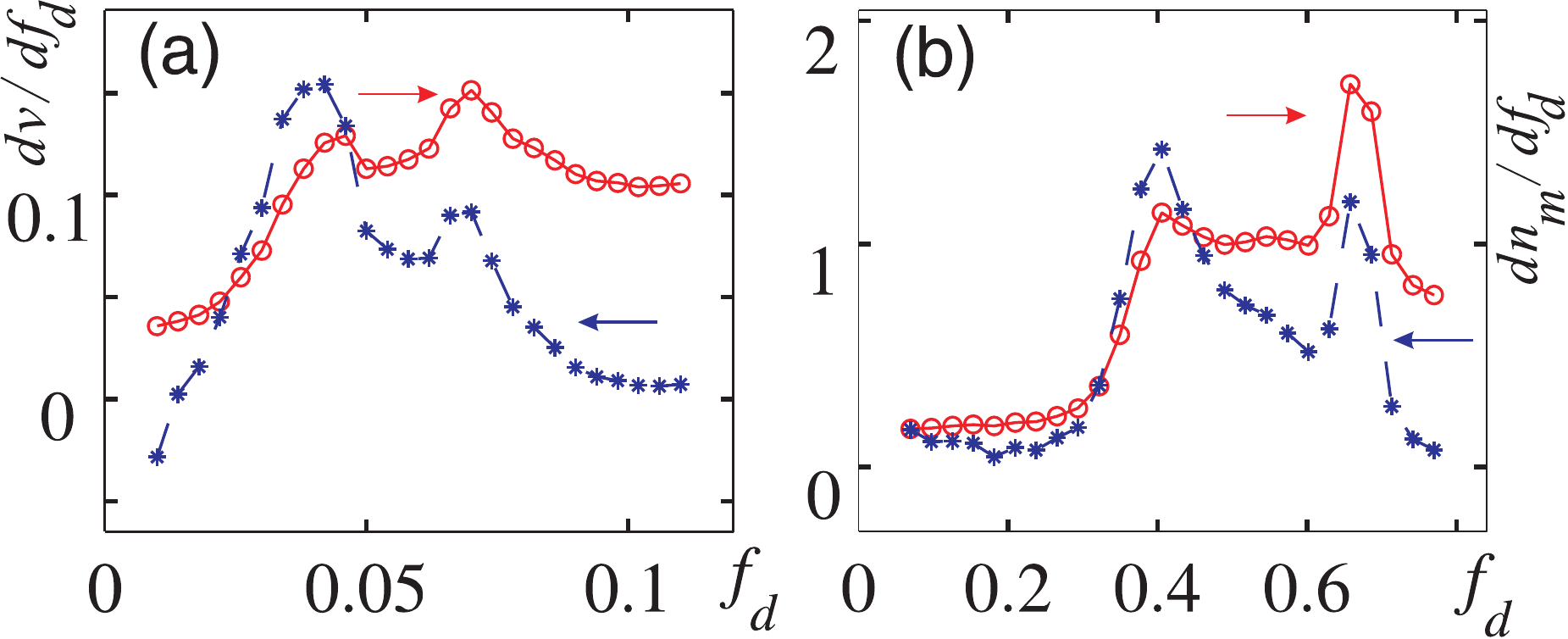} 
  \caption{The derivative of the average velocity $dv/df_{d}$ (solid red line and circles) and of the number of mobile particles $dn/df$ (dashed blue line and stars) vs. $f_d$, 
for $f_p=$ (a) 0.3; (b) 0.9. 
}
\label{fig:dv_plus_dn}
\end{figure}

\section{Phase diagram} 

The dynamical phase diagram 
in the plane of 
``pinning strength $f_p$ versus driving force $f_d$'' 
is shown in Fig.~\ref{fig:phase}. 
The pinned state is formed when the driving force is 
extremely small, i.e., 
insufficient to destroy the patterns. 
The depinned state corresponds to large driving forces. 
Note that due to collective depinning, assisted by the elastic force between the particles, the actual threshold is smaller than the maximum pinning force (dashed line).
For weak pinning, there are two dynamical regimes: 
pinned clusters (labyrinths) with particle flow inside the labyrinth (Fig.~\ref{fig:fp_small}(a)) and 
depinned clusters (Fig.~\ref{fig:fp_small}(b)) or stripes. 
For intermediate pinning 
we distinguish: 
(i) ``pinned clusters with chains-like flow''; 
(ii) ``intermediate regime'' including 
chains-like flow with a background of disordered pinned particles (Fig.~\ref{fp_intermediate}(a)), 
reordered depinned clusters (Fig.~\ref{fp_intermediate}(b)), 
and oriented stripes (Fig.~\ref{fp_intermediate}(c)); 
and 
(iii) ``depinned elastic flow'' (Fig.~\ref{fp_intermediate}(d)). 
With increasing 
$f_{p}$, 
the driving range for the formation of depinned clusters and stripes becomes very narrow.

\subsection{Peak effect} 

To analyze the curvature of the $<v>-f_d$-curve 
(Fig.~\ref{fig:fv}), 
we take the derivative $d<v>/df_{d}$ and find that each curve displays {\it two} main peaks 
(Fig.~\ref{fig:dv_plus_dn}). 
With increasing 
$f_{d}$, 
there are two contributions to $<v>$: 
the varying number of mobile particles and 
the increasing particle velocity. 
The nonlinear behavior of $d<v>/df_{d}$ is due to the former contribution. 
The variation of the number of mobile particles, $n_m=v/f_d$, is: 
$ 
{dn_m}/{df_d}=({dv}/{df_d})f_d^{-1}-vf_d^{-2}. 
$ 
As shown in Fig.~\ref{fig:dv_plus_dn}, ${dn_m}/{df_d}$ also has two peaks, which are 
related to the peaks in $dv/df$. 
The two-peak structure in ${dn_m}/{df_d}$ (and $d<v>/df_{d}$) 
is explained as follows. 
The first peak ($f_d\approx0.5f_p$): 
when $f_{d}$ increases, it depins more particles resulting in the
growth of $dn_m$, then the number of pinned particles decreases. 
The second peak appears when $f_d$ reaches the depinning threshold. 

The distinct dynamical patterns can also be characterized by 
other ``dynamical order parameters'' (for static patterns, 
see Ref.~\cite{Zhao2012a}). 
The transitions (crossovers) between the dynamical patterns 
are observed as features in these order parameters, 
such as the velocity distribution and the surface-area-to-volume ratio. 
(Note that the transitions between the different dynamical phases are rather smooth, although the different phases are well-distinguished as peaks in the corresponding order parameters.)

\subsection{Velocity distribution} 

In Fig.~8, we plot the dynamical patterns and the corresponding velocity distributions for pinning strength $f_p=0.5$. 
For driving force $f_d<0.21$ (see Fig.~8(a)), there is only one peak positioned at $0$ in the plot ``probability $p(v)$ versus velocity $v$''. 
This corresponds to the state when all particles are pinned. 

The second peak, which starts to develop as a weak feature at applied driving $f_d\geq0.21$, corresponds to the situation when the mobilized (depinned) particles start to form clusters (see Figs.~8(b)). 
Further increasing $f_d$, clearly leads to the formation of clusters, which strengthens the second peak. 
Simultaneously, the first peak weakens (see Figs.~8(c)). 
The two-peak structure shown in Figs.~8(c) corresponds to moving clusters (second peak) and pinned ones (first peak). 

Stripes are formed when the second peak is strong enough, i.e., enough mobilized (unpinned) particles are generated to form stripes (see Figs.~8(d)). 
Finally, when $f_d>f_p$, all the particles are mobilized. 
As a result, the first peak disappears, and the second peak shifts towards higher values indicating that all the particles participate in the motion, with the same velocity.

\begin{figure} 
    \includegraphics[width=8.0cm]{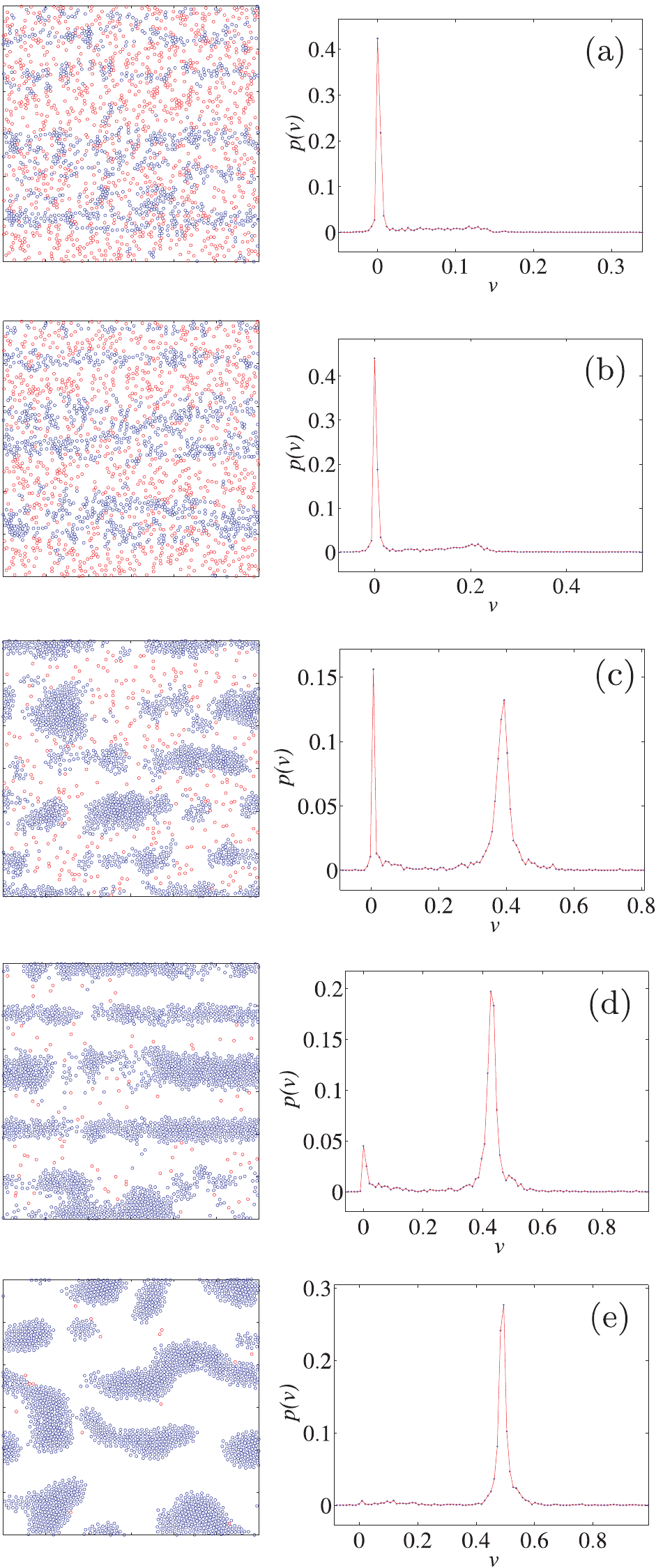} 
    \caption{Left panel: Dynamical patterns for pinning strength 
    $f_p=0.5$ and driving force $f_d=$ (a)0.13; (b)0.21; (c)0.39; (d)0.43; (e)0.49. The mobile particles are shown by blue (grey) dots, while the pinned or collectively pinnned particles are shown by red (black) dots. Right panel: Corresponding velocity distributions $p(v)$. }
\label{fp_intermediate}
\end{figure} 

\subsection{Surface-area-to-volume ratio} 

In Fig.~9, we plot ``surface-area-to-volume ratio $\gamma_{sv}$ versus driving force $f_d$''. 
(Note that the state with all the particles pinned is not considered in this section.) 
For weak pinning ($f_p=0.1$), there are only two dynamical regimes as discussed in the main text (see Fig.~9(a)). 
As $f_d$ increases, $\gamma_{sv}$ first increases, since the applied driving destroys the preexisted static patterns, i.e., labyrinths or clusters. 
The fast decrease of the function $\gamma_{sv}$ corresponds to the onset of the collective depinning of clusters. 
The saturation region for $f_d\geq0.074<f_p$ indicates that the threshold of collective depinning in this case is {\it much lower} than the pinning force. 

For intermediate pinning ($0.3\leq f_p \leq 0.7$), similarly to the weak pinning case, $\gamma_{sv}$ first increases when increasing $f_d$ (see Figs.~9(b) and (c)). 
Then $\gamma_{sv}$ rapidly decreases when mobilized particles start to clustering. 
Further increase of $f_d$ results, on the one hand, in depinning of  additional individual particles (they are counted as ``surface particles''). 
On the other hand, small clusters still can survive for these values of driving force $f_d$ (smaller clusters correspond to larger $\gamma_{sv}$). 
As a result of the competition of these two effects, a weak plateau (see Figs.~9(b) and (c)) appears in $\gamma_{sv}$. 
The latter effect disappears when all small clusters merge into stripes, thus $\gamma_{sv}$ start to rapidly decrease again. 
A fast drop in $\gamma_{sv}$ appears when $f_d>f_p$. 
The reason is that in this case, the particles form large clusters which have much lower $\gamma_{sv}$ than stripes. 

For strong pinning case ($f_p\geq0.9$), the dynamical pattern change  directly from a disordered flow to depinned ordered clusters. 
Thus a fast decay of $\gamma_{sv}$ is observed when $f_d$ is close to $f_p$. 

Note that for all the values of pinning $f_p$ considered above, the function $\gamma_{sv}$ has a constant minimum value ($\approx0.2$), when $f_d>f_p$, which indicates that they all share similar states (depinned ordered clusters).

\begin{figure*}
    \includegraphics[width=12cm]{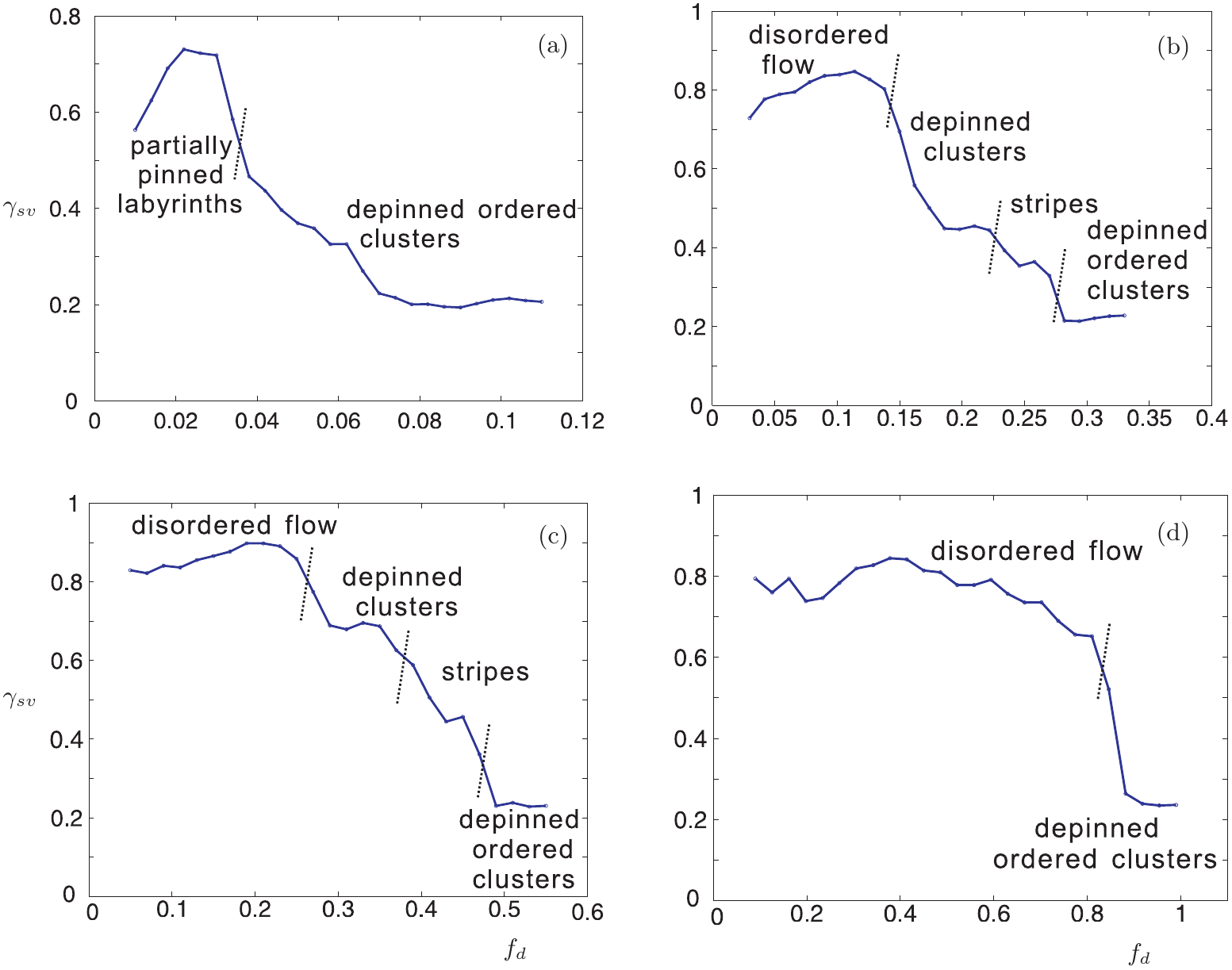} 
    \caption{Surface-area-to-volume ratio $\gamma_{sv}$ versus driving force $f_d$, for $f_p=0.1$ (a); $f_p=0.3$ (b); $f_p=0.5$ (c); and $f_p=0.9$ (d).  } 
\label{fp_SAVR} 
\end{figure*}

\section{Conclusions} 

Analyzing the dynamics of patterns formed due to a short-range non-monotonic interaction, 
we found, 
depending on the strength of the pinning, 
a variety 
of dynamical regimes 
including: 
a disordered motion 
(when the pre-existed static patterns are destroyed), 
the motion of depinned clusters, and 
the formation of stripes. 
Strikingly, the stripes 
can be either longitudinal 
(i.e., formed in the direction of the driving force) 
or 
transverse 
(i.e., formed in the direction normal to the driving), 
depending on the pinning regime: weak or strong, correspondingly.
We also found that, 
unlike in the 
previously studied 
static case, 
the motion on a random substrate dynamically selects the optimal cluster size.
Clusters smaller than this size, lose their particles due to the effect of friction and finally ``evaporate'', while large clusters collect these individual particles and grow untill only large enough clusters remain in the system.
The mechanism of this dynamical size selection effect is similar to the critical wettability of a water droplet moving on either an over-wetted or a dry substrate.
We found a pronounced double-peak structure 
in the variation of the $<v> - f_{d}$-curve 
which is explained by: 
(i) increasing the number of depinned particles and the simultaneous decrease of the number of pinned particles which can be depinned, and 
(ii) approaching the depinning threshold for individual particles.

Our findings can be useful for the analysis of dynamical pattern formation in various systems with competing range interaction including colloids, vortices in superconductors, etc., as well as 
for the deeper understanding of wetting phenomena and microscopic friction. 
In particular, we believe that the dynamical patterns, predicted in our work, can be verified in experiments with driven colloids and with driven vortices in two-band superconductors such as MgB$_{2}$.

\section{Acknowledgment} 

This work was supported by the ``Odysseus'' Program of the Flemish Government and the Flemish Science Foundation (FWO-Vl).


\section{Supplemental Material: Particle distributions} 

Here we show examples of snapshots illustrating various dynamical regimes and transitions (crossovers) between them, for $f_{p} = 0.7$ and varying driving force $f_{d}$. 
In particular, the crossower from disordered flow to clusters is shown in Fig.~10, and the crossower from stripes to clusters is presented in Fig.~11. 


\begin{figure}
\includegraphics[width=7.5cm]{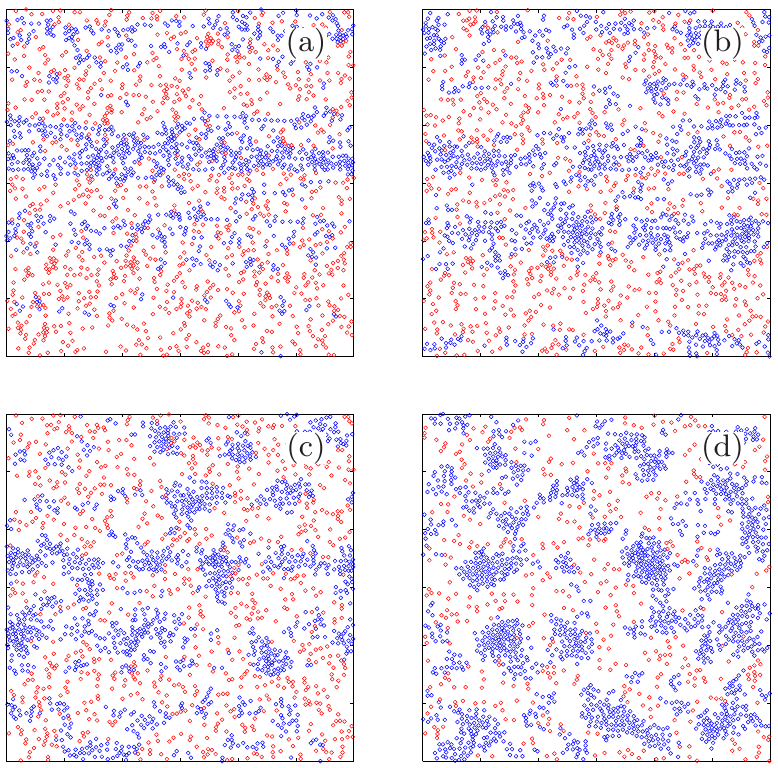}
\vspace{0.5cm}
\caption{
The crossover from disordered flow to clusters for $f_p=0.7$ and $f_d=0.21$ (a); $0.378$ (b); $0.406$ (c); $0.49$ (d).} 
\end{figure}



\begin{figure}
\includegraphics[width=7.5cm]{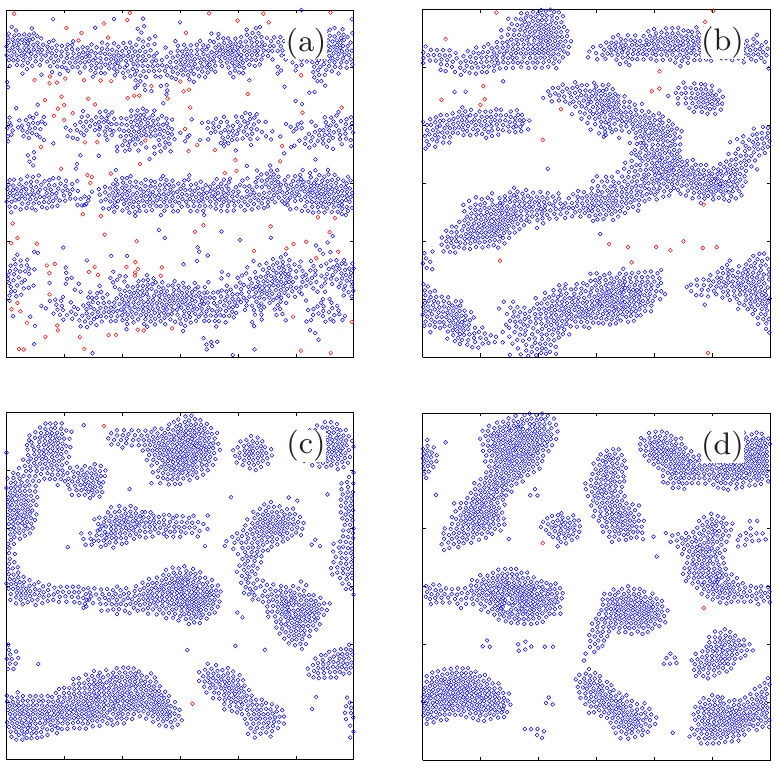} 
\vspace{0.5cm}
\caption{The crossover from stripes to clusters for $f_p=0.7$ and $f_d=0.63$ (a); $0.686$ (b); $0.714$ (c); $0.742$ (d).} 
\end{figure}


\begin{thebibliography}{10}

\bibitem{Seul1995}
M. Seul and D. Andelman, Science {\bf 267},  476  (1995).

\bibitem{Reichhardt2003}
C. Reichhardt, C.~J. Olson, I. Martin and A.~R. Bishop, Europhys. Lett. {\bf
  61},  221  (2003).

\bibitem{Reichhardt2003a}
C. Reichhardt, C.~J.~O. Reichhardt, I. Martin and A.~R. Bishop, Phys. Rev.
  Lett. {\bf 90},  026401  (2003).

\bibitem{Nelissen2005}
K. Nelissen, B. Partoens and F.~M. Peeters, Phys. Rev. E {\bf 71},  066204
  (2005).

\bibitem{Vedmedenko2007}
E.~Y. Vedmedenko, {\em Competing interactions and patterns in nanoworld}
  (WILEY-VCH Verlag GmbH \& Co. KGaA, 2007).

\bibitem{Zhao2012a}
H.~J. Zhao, V.~R. Misko and F.~M. Peeters, New Journal of Physics {\bf 14},
  063032  (2012).

\bibitem{Whitesides2002}
G. Whitesides and M. Boncheva, Proc. Natl. Acad. Sci. U. S. A. {\bf 99},  4769
  (2002).

\bibitem{Ball1999}
P. Ball, {\em The Self-Made Tapestry: Pattern Formation in Nature} (Oxford
  Univ. Press, Oxford, U.K., 1999).

\bibitem{Reichhardt2002a}
C. Reichhardt and C.~J. Olson, Phys. Rev. Lett. {\bf 89},  078301  (2002).

\bibitem{Cao2003}
Y. Cao, J. Chen, Y. Wang, Z. Jiao and W.~K. Chow, Phys. Rev. B {\bf 68},
  134209  (2003).


\bibitem{Pertsinidis2008}
A. Pertsinidis and X.~S. Ling, Phys. Rev. Lett. {\bf 100},  028303  (2008).

\bibitem{Bohlein2012}
T. Bohlein, J. Mikhael and C. Bechinger, Nature Materials {\bf 11},  126
  (2012).

\bibitem{Gruner1988}
G. Gr\"uner, Rev. Mod. Phys. {\bf 60},  1129  (1988).

\bibitem{Balents1995}
L. Balents and M.~P.~A. Fisher, Phys. Rev. Lett. {\bf 75},  4270  (1995).

\bibitem{Reichhardt2001a}
C. Reichhardt, C.~J. Olson, N. Gr\o{}nbech-Jensen and F. Nori, Phys. Rev. Lett.
  {\bf 86},  4354  (2001).

\bibitem{Bhattacharya1993}
S. Bhattacharya and M.~J. Higgins, Phys. Rev. Lett. {\bf 70},  2617  (1993).

\bibitem{Yaron1995}
U. Yaron, Nature {\bf 376},  753  (1995).

\bibitem{Duarte1996}
A. Duarte, Phys. Rev. B {\bf 53},  11336  (1996).

\bibitem{Pardo1996}
F. Pardo, Phys. Rev. Lett. {\bf 78},  4633  (1996).

\bibitem{Pardo1998}
F. Pardo, F. de~la Cruz, P.~L. Gammel, E. Bucher and D.~J. Bishop, Nature {\bf
  396},  348  (1998).

\bibitem{Koshelev1994}
A.~E. Koshelev and V.~M. Vinokur, Phys. Rev. Lett. {\bf 73},  3580  (1994).

\bibitem{Moon1996}
K. Moon, T. Scalettar and T. Gergely, Phys. Rev. Lett. {\bf 77},  2778  (1996).

\bibitem{Giamarchi1996}
T. Giamarchi and P. Le~Doussal, Phys. Rev. Lett. {\bf 76},  3408  (1996).


\bibitem{Balents1997}
L. Balents, M.~C. Marchetti and L. Radzihovsky, Phys. Rev. Lett. {\bf 78},  751
   (1997).

\bibitem{Olson1998}
C.~J. Olson, C. Reichhardt and F. Nori, Phys. Rev. Lett. {\bf 81},  3757
  (1998).

\bibitem{Rosenstein2010}
B. Rosenstein and D. Li, Rev. Mod. Phys. {\bf 82},  109  (2010).

\bibitem{Pogosov2010}
W.~V. Pogosov, H.~J. Zhao, V.~R. Misko and F.~M. Peeters, Phys. Rev. B {\bf 81},  024513  (2010).

\bibitem{Reichhardt1997}
C. Reichhardt, C.~J. Olson and F. Nori, Phys. Rev. Lett. {\bf 78},  2648
  (1997).


\bibitem{Misko2006b}
V.~R. Misko, S. Savel'ev, A.~L. Rakhmanov and F. Nori, Phys. Rev. Lett. {\bf
  96},  127004  (2006).

\bibitem{Misko2007a}
V.~R. Misko, S. Savel'ev, A.~L. Rakhmanov and F. Nori, Phys. Rev. B {\bf 75},
  024509  (2007).

\bibitem{Misko2005}
V. Misko, S. Savel'ev and F. Nori, Phys. Rev. Lett. {\bf 95},  177007  (2005).

\bibitem{Misko2006}
V.~R. Misko, S. Savel'ev and F. Nori, Phys. Rev. B {\bf 74},  024522  (2006).

\bibitem{Villegas2006}
J.~E. Villegas, M.~I. Montero, C.-P. Li and I.~K. Schuller, Phys. Rev. Lett.
  {\bf 97},  027002  (2006).

\bibitem{Kemmler2006}
M. Kemmler, C. G\"urlich, A. Sterck, H. P\"ohler, M. Neuhaus, M. Siegel, R.
  Kleiner and D. Koelle, Phys. Rev. Lett. {\bf 97},  147003  (2006).

\bibitem{Silhanek2006}
A.~V. Silhanek, W. Gillijns, V.~V. Moshchalkov, B.~Y. Zhu, J. Moonens and
  L.~H.~A. Leunissen, Appl. Phys. Lett. {\bf 89},  152507  (2006).

\bibitem{Misko2010}
V.~R. Misko, D. Bothner, M. Kemmler, R. Kleiner, D. Koelle, F.~M. Peeters and
  F. Nori, Phys. Rev. B {\bf 82},  184512  (2010).

\bibitem{Xu2011b}
X.~B. Xu, H. Fangohr, Z.~H. Wang, M. Gu, S.~L. Liu, D.~Q. Shi and S.~X. Dou,
  Phys. Rev. B {\bf 84},  014515  (2011).

\bibitem{OlsonReichhardt2011}
C.~J. Olson~Reichhardt, C. Reichhardt and A.~R. Bishop, Phys. Rev. E {\bf 83},
  041501  (2011).


\bibitem{Brandt2011}
E. Brandt and M. Das, J. Supercond. Novel Magn. {\bf 24},  57  (2011).

\bibitem{Xu2011}
X.~B. Xu, H. Fangohr, S.~Y. Ding, F. Zhou, X.~N. Xu, Z.~H. Wang, M. Gu, D.~Q.
  Shi and S.~X. Dou, Phys. Rev. B {\bf 83},  014501  (2011).

\bibitem{Moshchalkov2009}
V. Moshchalkov, M. Menghini, T. Nishio, Q.~H. Chen, A.~V. Silhanek, V.~H. Dao,
  L.~F. Chibotaru, N.~D. Zhigadlo and J. Karpinski, Phys. Rev. Lett. {\bf 102},
   117001  (2009).

\bibitem{Nishio2010}
T. Nishio, V.~H. Dao, Q. Chen, L.~F. Chibotaru, K. Kadowaki and V.~V.
  Moshchalkov, Phys. Rev. B {\bf 81},  020506  (2010).

\bibitem{Babaev2005}
E. Babaev and M. Speight, Phys. Rev. B {\bf 72},  180502  (2005).

\bibitem{Babaev2010}
E. Babaev, J. Carlstr\"om and M. Speight, Phys. Rev. Lett. {\bf 105},  067003
  (2010).

\bibitem{Chaves2011}
A. Chaves, L. Komendov\'a, M.~V. Milo\ifmmode \check{s}\else
  \v{s}\fi{}evi\ifmmode~\acute{c}\else \'{c}\fi{}, J.~S. Andrade, G.~A. Farias
  and F.~M. Peeters, Phys. Rev. B {\bf 83},  214523  (2011).

\bibitem{Komendova2011}
L. Komendov\'a, M.~V. Milo\ifmmode \check{s}\else
  \v{s}\fi{}evi\ifmmode~\acute{c}\else \'{c}\fi{}, A.~A. Shanenko and F.~M.
  Peeters, Phys. Rev. B {\bf 84},  064522  (2011).

\bibitem{Lin2011}
S.-Z. Lin and X. Hu, Phys. Rev. B {\bf 84},  214505  (2011).

\bibitem{Komendova2012}
L. Komendov\'a, Y. Chen, A.~A. Shanenko, M.~V. Milo\ifmmode \check{s}\else
  \v{s}\fi{}evi\ifmmode~\acute{c}\else \'{c}\fi{} and F.~M. Peeters, Phys. Rev.
  Lett. {\bf 108},  207002  (2012).

\bibitem{Silaev2012}
M. Silaev and E. Babaev, Phys. Rev. B {\bf 85},  134514  (2012).

\bibitem{Gutierrez2012}
J. Gutierrez, B. Raes, A.~V. Silhanek, L.~J. Li, N.~D. Zhigadlo, J. Karpinski,
  J. Tempere and V.~V. Moshchalkov, Phys. Rev. B {\bf 85},  094511  (2012).

\bibitem{Reichhardt2005}
C. Reichhardt, C.~J.~O. Reichhardt and A.~R. Bishop, Europhys. Lett. {\bf 72},
  444  (2005).

\bibitem{Zhao2012}
H.~J. Zhao, V.~R. Misko and F.~M. Peeters, Physica C: Superconductivity {\bf
  479},  130   (2012).

\bibitem{Reichhardt2012a}
C. Reichhardt, J. Drocco, C.~O. Reichhardt and A. Bishop, Physica C:
  Superconductivity {\bf 479},  15   (2012).

\bibitem{pbc} 
To check the effect of boundaries, an enlarged simulation 
box was used, namely, we tried $L_{x} = L_{y} = L = 240$ 
and $L_{x} = L_{y} = L = 360$ which revealed no change in 
the resulted patterns. 

\bibitem{cutoff} 
Since both terms of the interaction force are represented by modified Bessel functions, which decay exponentially for large distances, the used cut-off procedure is safe: our trial simulations with a cut-off of $r>16\lambda$ did not reveal any impact on the results. 

\bibitem{temp} 
Nonzero temperature would result in broadening of the particle  trajectories and ultimately in destruction of dynamical patterns. 


\bibitem{Gennes1985}
P.~G. de~Gennes, Rev. Mod. Phys. {\bf 57},  827  (1985).

\bibitem{Bonn2009}
D. Bonn, J. Eggers, J. Indekeu, J. Meunier and E. Rolley, Rev. Mod. Phys. {\bf
  81},  739  (2009).

\bibitem{Malescio2003}
G. Malescio and G. Pellicane, Nat. Mater. {\bf 2},  97  (2003).




\end{thebibliography}
\end{document}